# Low frequency electrical waves in ensembles of proteinoid microspheres

Panagiotis Mougkogiannis*, Andrew Adamatzky

*Unconventional Computing Laboratory, UWE, Bristol, UK*

**Abstract**

Proteinoids (thermal proteins) are produced by heating amino acids to their melting point and initiation of polymerisation to produce polymeric chains. Amino acid-like molecules, or proteinoids, can condense at high temperatures to create aggregation structures called proteinoid microspheres, which have been reported to exhibit strong electrical oscillations. When the amino acids LGlutamic acid (L-Glu) and L-Aspartic acid (L-Asp) were combined with electric fields of varying frequencies and intensities, electrical activity resulted. We recorded electrical activity of the proteinoid microspheres' ensembles via a pair of differential electrodes. This is analogous to extracellular recording in physiology or EEG in neuroscience but at micro-level. We discovered that the ensembles produce spikes of electrical potential, an average duration of each spike is 26 min and average amplitude is 1 mV. The spikes are typically grouped in trains of two spikes. The electrical activity of the ensembles can be tuned by external stimulation because ensembles of proteinoid microspheres can generate and propagate electrical activity when exposed to electric fields.

*Keywords:* thermal proteins, proteinoids, microspheres, unconventional computing

## 1. Introduction

Thermal proteins (proteinoids) [1] are produced by heating amino acids to their melting point and initiation of polymerisation to produce polymeric chains. In aqueous solution proteinoids swell forming microsphere [1]. The proteinoid microspheres have been considered as proto-neurons [2, 3, 4]. Emerging cells had similar to proteinoid microspheres' structure. Artificial fossilization of the laboratory products closely resemble ancient microfossils from ancient strata [2].

The findings on propagation of excitation waves in ensembles of proteinoid microspheres will open ways for studying coupling of biochemical and other phenomena in ways that cannot be accomplished with other models such as bilayer membranes. The proteinoids are sufficiently accurate prototypes of terrestrial protocells with bioelectrical properties [5, 6].

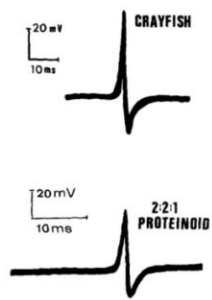

Figure 1: A comparison of two distinct electrical signal types is illustrated in the figure. The first form of nerve impulse is the action potential of a crayfish stretch receptor neuron. These neurons are specialised to sense changes in either length or pressure. Proteinoid microspheres, a form of artificial cell created by heat polymerisation of amino acids, generate the second electrical signal. Taken from [2].

Having most characteristics of excitable cells proteinoid microspheres have been considered as protoneurons [7]. By studying proteinoid oscillations, we can learn more about their molecular dynamics and how they interact with one another [8]. This information is essential for the search for alien life and for understanding how life first developed on Earth [9].

In a 1985 paper published in the journal *Biosystems*, Przybylski et al. [7] proposed the concept of electrical oscillations in proteinoid cells. Although they originate in different cell types, the action potentials of crayfish stretch receptor neurons and proteinoid microspheres are very similar (Fig.1).

In a paper by Margel et al., glutamic acid was found to be essential for the proper folding of proteinoids [10]. Glutamic acid is one of the most important amino acids in the polymerization of proteinoids. It acts as a solvent and condensation agent, and is also a key component of pyroglutamic acid, which is essential for cross-linking proteinoids. Glutamic acid is also involved in the formation of disulfide bonds [11, 12] which are essential for the stability of proteinoids. The process of self-assembly is a key factor in the formation of both polystyrene formation [13, 14] and proteinoids [15, 16]. However, there are some key differences in the way the two materials form. Polystyrene form via a process of self-assembly in the presence of surfactants. The surfactants help to lower the surface tension of the polystyrene, allowing to self assembly into nanoparticles. In contrast, proteinoids form through self assembly in the absence of surfactants. Instead the amino acids that make up proteinoids are able to self assemble due to their hydrophobic and hydrophilic interactions[17].

A consensus was achieved that proteinoids directly led into neurons, which then self-associated into brains [1]. Thus ensembles of proteinoid microspheres can be seen as proto-brains, potentially, capable for sensorial fusion, information transmission and processing. In [18] we proposed to design and prototype unconventional computing devices from ensembles of proteinoid microspheres. There are two primary tenets upon which this idea rests. The first method borrows ideas and prototypes from excitable medium computers, such as their algorithms. To simulate the self-organising behaviour of living systems, scientists turn to computers based on excitable media. This method paves the way for a dynamic and adaptable structure that can quickly respond to shifting requirements. As for the second, it makes use of reservoir computing methods. The machine learning approach known as reservoir computing makes use of a huge pool of neurons with predetermined connections that may be easily tweaked to process new types of data. These computers have the potential to transform the way we handle data and address difficult problems by integrating the two major methodologies. Moreover, this idea may have far-reaching consequences for the future of artificial intelligence.

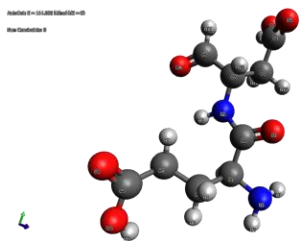

Figure 2: The ionised carboxyl and amino groups of an L-Glu:L-Asp polypeptide are depicted here, illustrating the zwitterionic structure of the peptide. Zwitterion molecules have a net electrical charge of zero due to the orientation of their polar groups [21].

All previous experimental studies on electrical activity of proteinoid microspheres a trans-membrane potential was recorded [19, 20, 7], this was alike to intra-cellular or patch-clamp techniques. In our studies we decided to test an analog of 'extra-cellular' recording by inserting a pair of differential electrodes into an ensemble of proteinoid micro-spheres. Our approach is somewhat similar to EEG recording of an integral neural activity: the electrical potential waves recorded from human brain are manifestations of the summarily activity of hundreds of thousands of neurons while the electrical activity of the proteinoid microspheres' ensembles is an integration of electrical discharges of thousands of proteinoids.

## 2. Methods

The preparation of the proteinoid L-Glu:L-Asp (Fig. 2) was achieved using L-Aspartic acid and L-Glutamic acid with reagent grade of greater than 98% and 99% , respectively. These compounds were purchased from Sigma-Aldrich and used without further purification. The proteinoid was prepared by combining the two amino acids in a 1:1 molar ratio to a magnetically stirred 35 ml vial. The reaction vessel was then heated to 290 degrees Celsius for two days before being cooled to room temperature.

At 80 degrees Celsius, 100 mg of the proteinoid was added to 10 mL of water in a vial. This made it possible for the proteinoid to dissolve in water. After the product had been thoroughly stirred, it was dialyed via a cellulose membrane (3500 Da MWCO). This allowed for the removal of any excess reagents, leaving the proteinoid in its purest form. The proteinoid solution was sealed and kept undisturbed for 12 to 24 hours to permit dialysis. The product was then collected once the membrane was removed.

An experimental approach was carried out to record the proteinoid voltage utilising a PicoLog USB data logger ADC 24 with a range of up to ± 2.5 V. A centimetre separated the platinum iridium electrodes, which were also connected to the data recorder. The one-second sampling interval was chosen. Mixing an adequate amount of proteinoid powder with distilled water while stirring until the powder was completely dissolved yielded the proteinoid solution. The solution was then drawn up with a 5 ml syringe and injected into the space between the two electrodes in a 5 ml vial.

The voltage was recorded in the input range to ± 130 mV and the sample rate was 1 second.

This work used the field emission scanning electron microscope (FESEM) FEI Quanta 650 to explore the proteinoids L-Glu:L-Asp.To acquire a precise SEM image of the proteinoid, samples were initially coated with a thin layer of gold. Gold plating supplies the electron beam with a conductive surface, resulting in a high-quality image. Additionally, the gold coating protects the sample from damage induced by the electron beam, which might result in sample degradation. The particle size distribution was estimated using the Image analysis software ImageJ [22].

## 3. Results

Spheroid clusters with a diameter of less than 100 nm may be seen forming in the Scanning Electron Microscopy (SEM) image of the proteinoid ensembles L-Glu:L-Asp (Fig. 3a). Agglomeration of proteinoid molecules leads to the production of these microspheres. The Lorentzian distribution of particle size and polydispersity of proteinoid nanoparticles is shown in (Fig. 3b). A polydispersity of 99.1 % was calculated from the fitted data, which indicated that the average particle diameter was 25 nm. This is evidence of remarkable consistency. In addition, a scanning electron micrograph (SEM) image of connected nanostructures is displayed in Fig. 3c. Nanoparticles of proteinoids are microscopic spheres that can aggregate into bigger structures. The Lorentzian distribution shows the nanoparticles' size and polydispersity and sheds light on how the particles clump together. When the majority of the particles fall to the left side of the curve, we say that the distribution is skewed to the left. In other words, little particles are more robust than big ones. What this means is that the median particle size is larger than the mean particle size. Particle size distributions typically have a left-skewed shape because small particles prefer to cluster together, increasing the proportion of small particles in the distribution. Larger particles are less likely to concentrate in one location since they are heavier and more difficult to shift.

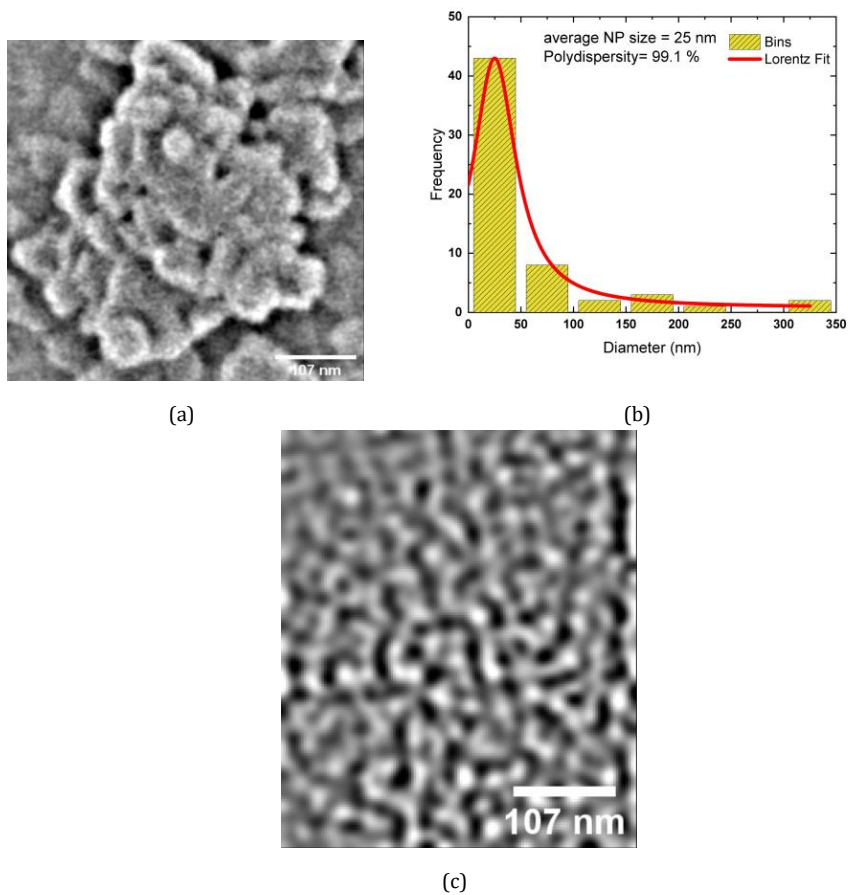

(a)                (b)

(c)

Figure 3: Characterisation of proteinoid ensembles. (a) SEM image shows the formation of proteinoid L-Glu:L-Asp nanoparticles, which are created by the self-assembly of amino acids and monomers. (b) In a Lorentzian fit to the SEM data, the nanoparticles' small diameter of 25 nm and polydispersity of 99.1 % are both indicative of a highly uniform size distribution. (c) This figure shows how the proteinoid's microspheresspheres are interconnected by a complex system of channels. Proteinoids rely on these channels for signal transmission and reception, allowing them to function properly.

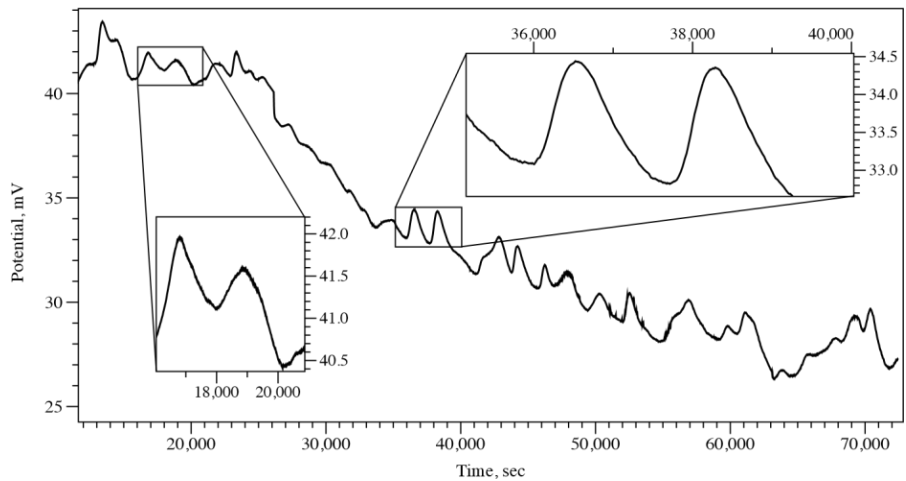

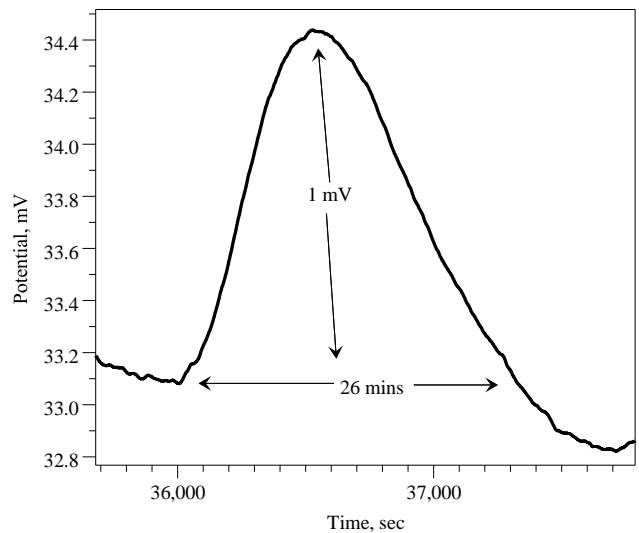

Figure 4: Electrical activity recorded in the ensemble of proteinoid microspheres via a pair of differential electrodes. (a) The activity recorded during 21 hours. Few characteristic voltage spikes are magnified in the inserts. (b) A typical spike of electrical potential.

Fig. 4a shows electrical potential, recorded in a pair of differential electrodes inserted in the ensemble of proteinoid microspheres, of the proteinoid microsphere ensembles made of L-Glu:L-Asp. Pronounced oscillation in the electrical output indicates that there is significant electrical activity between the microspheres. A typical spike of electrical potential is shown in Fig. 4b.

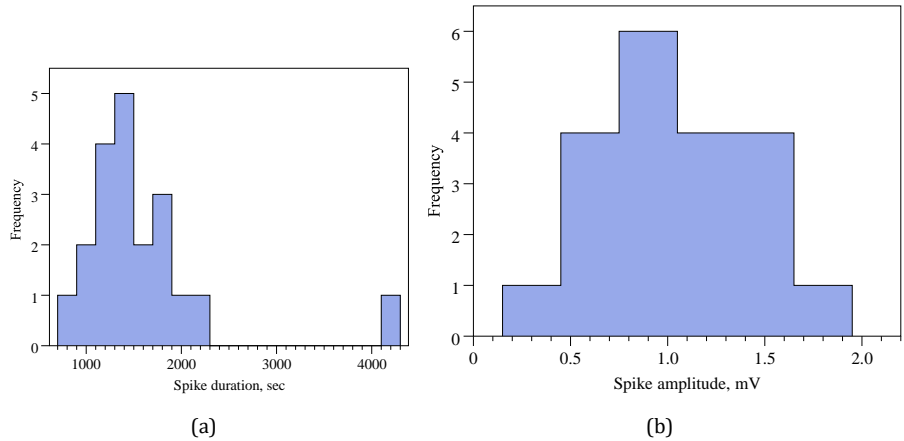

Figure 5: Characteristic of spiking. (a) Distribution of spike lengths, bin size is 200 sec. (b) Distribution of spike amplitude, bin size is 0.3 mV.

Of twenty spikes measured we found that average spike duration is 1574 sec, $\sigma$ = 697, median 1360 sec, minimum 830 sec, maximum 4130 sec. That is a typical spike recorded lasts for c. 26 min. Spike duration distribution is show in Fig. 5a. The distribution has relatively short right tail but still more spread to the right, as skewness $\gamma_1$ =2.79568212 and kurtosis $\beta_2$=13.2915541.

Average spike amplitudes varied from minimum 0.3 mV to maximum 1.9 mV, average amplitude 1 mV, $\sigma$ = 0.4, median 1 mV. Distribution of spike amplitudes is shown in Fig. 5b. The distribution is almost symmetric with very small leaning to the right: skewness $\gamma_1$ =0.052873856 and kurtosis $\beta_2$=2.80547952.

We say spikes form a train when a distance between consecutive spikes does not exceed a double of average spike width. All trains of spikes observed have just two spikes each. An example of a train of spikes is shown in magnifying in insert at the top right of the plot Fig. 4a.

Ensembles of proteinoid microspheres, made of L-Glu:L-Asp, can be stimulated to create electrical signals of varying amplitudes and frequencies by being subjected to a range of frequencies, from 0.001 Hz to 100 kHz, for 1 minute. Exposed to these frequencies, the proteinoid will begin to generate electrical oscillations with a period of 1 hour and voltages ranging from -2.29 mV to 12.771 mV. After 1 minute of stimulation at frequencies ranging from 0.001 Hz to 100 kHz, the electrical oscillations of a proteinoid containing L-Glu:L-Asp are depicted in Fig. 6. This suggests that the proteinoid can be stimulated by electrical oscillations of varying frequencies. Therefore, it appears that the proteinoid can pick up on the frequencies and react appropriately. These findings add to the growing body of evidence that proteinoids can sense and react to their environment.

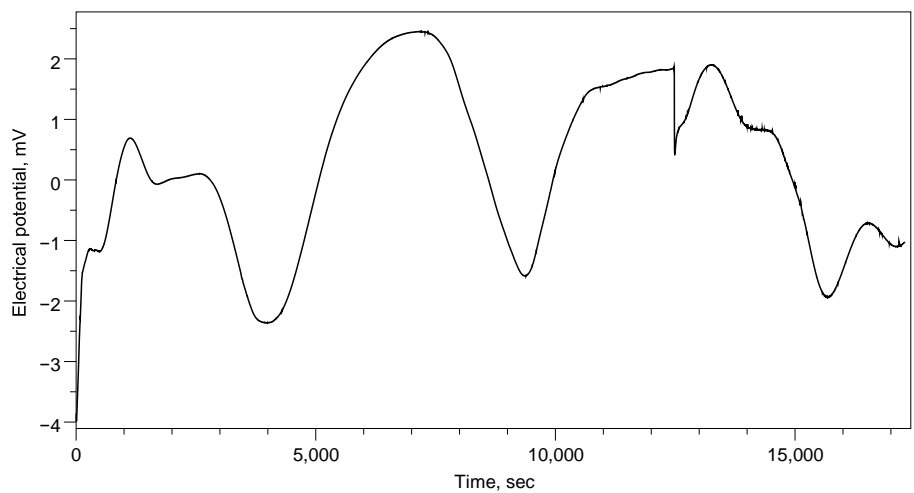

Figure 6: The pronounced electrical oscillations of the L-Glu:L-Asp proteinoid microspheres ensembles after being stimulated for one min with frequencies between 0.001 Hz to 100 kHz.

## 4. Discussion

In laboratory experiments we demonstrated that ensembles of L-Glu:L-Asp proteinoid microspheres generated oscillations of electrical potential recorded via pair of differential electrodes. The spikes of the electrical potential reflect propagation of electrical activity in proteinoid ensembles. An average duration of a voltage spike is 26 min and average amplitude 1 mV. The value is much smaller than electrical potential amplitudes, ranging from 20 mV to 70 mV, recorded 'intra-spherically', between interior and exterior of proteinoid microspheres [19], this is because strength of a signal is substantially smaller in case of 'extra-spherical' recording. We can speculate that the spiking recorded is manifestation of the coordinated electrical activity propagating in the microspheres' ensembles. Reported duration of 'intra-spherically' recorded voltage spikes are around 1 sec. Distance between electrodes in our experiments was 10 mm. Media duration of spikes recorded in our experiments is 1360 sec. Assuming the activity propagates along a shortest path connecting the electrodes we can propose that there are c. 1360 proteinoid microspheres along the path, which might give an indicated diameter of a microsphere as c. 7 $\mu$m. This is less than the sizes of the microspheres, 20 $\mu$m to 200 $\mu$m [1], reported previously. Therefore we can also speculate that is might take some time, at least 3 sec, exact value to be investigated in further experiments, for electrical activity to propagation between the proteinoid microspheres.

Why do ensembles of proteinoid microspheres form?

Electrostatic interactions between the charged amino acids in the proteinoid are most likely responsible for the clustering. Since proteinoids are made up of amino acids, which are the foundation of proteins, their formation is significant. Proteinoids are able to organise themselves into a more sophisticated and structured structure by aggregating into nanoparticles. This higher level of complexity has the potential to improve the nanostructure's stability, which in turn could have future applications. The creation of proteinoid microspheres can be broken down into three distinct stages, each of which is seen in the image (Fig. 7) below. First, proteinoids cluster together by a process called nucleation. Next, the molecules will cluster together to create a larger aggregate. The clumps eventually coalesce into microscopic spheres. Below, each of these steps will be discussed in detail.

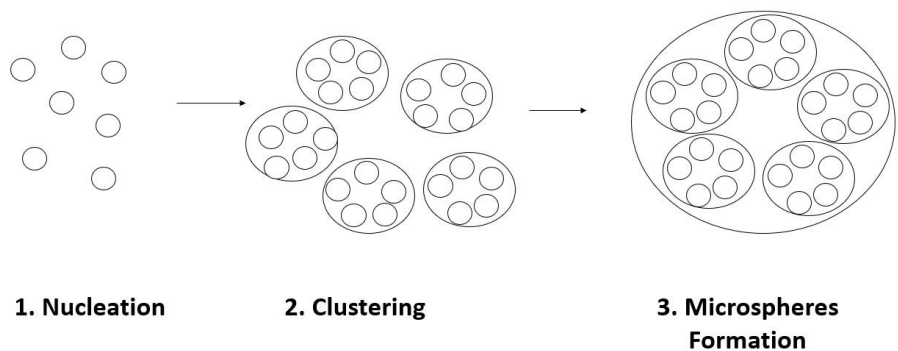

**1. Nucleation**     **2. Clustering**     **3. Microspheres Formation**

Figure 7: The three-step production process of proteinoid microspheres: nucleation, clustering, formation of microspheres.

1. Nucleation: The amino acid building blocks of proteins, called proteinoids, are heated in a watery solution. Due to this, 1-2 nm sized nano-spheres are formed.
2. Clustering: Following this, the nanospheres coalesce into bigger clusters of 50 nm in size. Hydrophobic interactions, van der Waals forces, and hydrogen bonding all play a role in facilitating this clustering.
3. Microspheres Formation: When additional molecules, including phospholipids and carbohydrates, are present in the solution, the clustered proteinoid molecules are further stabilised and form larger microspheres. This procedure produces microspheres of proteinoid several hundred nanometers in diameter that are extremely stable.

What are possible mechanisms of the oscillations?

The interaction between L-Glu and L-Asp is likely responsible for the fluctuations in electrical activity, which in turn imply a dynamic exchange of ions between the microspheres. L-Glu has a positive charge due to the partial positive charge on the carboxyl group and the partial negative charge on the side chain, whereas L-Asp has a partial negative charge on the carboxyl group and a partial positive charge on the side chain. This results in an electrostatic relationship between the two amino acids, which results in an electrical current and electrical oscillations. The presence of water molecules in the proteinoid solution further enhances the oscillations. The water molecules serve as electrolytes, so facilitating the flow of electrical current and enhancing the oscillations. The inclusion of other amino acids in the solution, such as L-Arginine, can further amplify the oscillations and amplify the electrical current.

Proteinoid aqueous solutions can be understood in terms of the behaviour of the proteinoid molecules, which causes electrical oscillations. Every molecule of a proteinoid in solution carries an electrical charge, which changes as the molecule shifts position and interacts with others. The proteinoids store and release electrical energy like a capacitor due to their changing charge. The solution's oscillatory behaviour is the result of the energy being released as electrical oscillations. Proteinoids can also form complex interactions with other solute molecules. Clusters of molecules can form as a result of this interaction, increasing the power of the electrical oscillation produced by the proteinoids. Clusters of proteinoid molecules contribute further to the solution's oscillatory behaviour by storing and releasing energy more efficiently than individual molecules.

Proteinoids interacting with other particles, such electrons, results in electrical oscillations. Charged proteinoids generate an electric field when they come into contact with electrons. The electrical waves originate from an underlying electrical field that oscillates at the same frequency as the proteinoids.

Proteinoids are sensitive to the frequencies of their environmental inputs, which impacts the period of their electrical oscillations. Proteinoids, which are made up of amino acid chains that interact with one another to form a polypeptide, may be electrically stimulated at frequencies ranging from 0.001 Hz to 100 kHz. Stimulating the proteinoid with these frequencies causes its period of electrical oscillations to grow from 25 minutes to a maximum of 2 hours and 45 minutes. The ability of the polypeptide to absorb and store energy from the environmental inputs is linked to the lengthening of the oscillation period. The polypeptide's structure becomes more ordered and stable when it is exposed to a frequency within its range and absorbs the energy, storing it in its bonds. Longer electrical oscillation periods come from the polypeptide's enhanced stability, which allows it to absorb and store more energy.

Further, the polypeptides' structure is connected to its sensitivity to particular frequencies. Amino acid sequences and protein structure are two factors that govern how proteinoids react to certain frequencies. Some proteinoids, for instance, respond better to high frequencies whereas others respond better to low frequencies. The period of oscillations expands as the frequency grows because the polypeptides' structure evolves to match it.

## 5. Conclusions

The electrical activity propagation in ensembles of proteinoid microspheres is a complicated process that depends on a wide range of variables, such as the kind and concentration of proteinoids, the resistivity of the medium, and the gap junctions between microspheres. Ion exchange between proteinoid microspheres and the conductive media is responsible for their neuron like spiking activity. The ensembles of proteinoid microspheres are demonstrated to be may serve as a model system for investigating the spread of electrical activity. While previous patch clamp and 'intra-microsphere' research demonstrated that proteinoids produce action-potential likes spikes, our research demonstrated it is possible to recorded EEG like pattern from the ensembles of proteinoid microspheres. Thus the proteinoid ensembles may be seen as nano-brains composed of hundred of neurons where each neuron is physically represented by a proteinoid microsphere.

**Acknowledgement**

The research was supported by EPSRC Grant EP/W010887/1 "Computing with proteinoids". Authors are grateful to David Paton for helping with SEM imaging and to Neil Phillips for helping with instruments.

**Data Availability**

All data generated or analysed during this study are included in this published article [and its supplementary information files].